\documentclass[twocolumn,showpacs,showkeys,preprintnumbers,aps]{revtex4-1}

\usepackage{graphicx}
\usepackage{graphics}
\usepackage{dcolumn}
\usepackage{bm}
\usepackage{epsfig}
\usepackage{longtable}
\usepackage{multirow}

\newlength\figwidth\figwidth=0.5\textwidth

\begin{document}

\title{New data on $0^+$ states in $^{158}$Gd}

\author {A.~I.~Levon$^{1}$, C.~Costache$^{3}$, T.~Faestermann$^{2}$, R.~Hertenberger$^{2}$,
A.~Ionescu$^{3,4}$, R.~Lica$^3$, A.~G.~Magner$^{1}$,  C.~Mihai$^{3}$, R.~Mihai$^{3}$, C.~R.~Nita$^{3}$,
S.~Pascu$^{3}$, K.~P.~Shevchenko$^{1}$, A.~A.~Shevchuk$^{1}$, A.~Turturica$^{3}$,  and H.-F.~Wirth$^{2}$}

\affiliation{$^1$ Institute for Nuclear Research, Academy of Science, Kiev, Ukraine}
\email[Electronic address: ] {alevon38@kinr.kiev.ua}

\affiliation{$^2$ Fakult\"at f\"ur Physik, Ludwig-Maximilians-Universit\"at M\"unchen, Garching, Germany}

\affiliation{$^3$ H.~Hulubei National Institute of Physics and Nuclear
Engineering, Bucharest, Romania}

\affiliation{$^4$ University of Bucharest,
Faculty of Physics,  Bucharest-Magurele, Romania}

\begin{abstract}
Excited states in the deformed nucleus $^{158}$Gd have been studied  in the (p,t)
reaction by using  the Munich Tandem and Q3D spectrograph. 30 new excited 0$^+$ states
(three tentative) have been assigned up to the 4.3 MeV excitation energy.
The total number of 34 excited  0$^+$ states (four tentatively
assigned) in a deformed nucleus,
close to a complete level scheme, offers a new opportunity to test nuclear models
and obtain more information on the structure of these special states.
\end{abstract}
\bigskip
\date{\today}


\maketitle

\section{Introduction}
Excited 0$^+$ states in nuclei   are   specific modes
 of nuclear excitations which were intensively studied,
 especially in the last few decades. They have a different structure  associated, e.g.,
 with pair vibrations, beta vibrations,  vibrations caused by spin-quadrupole forces,
 one- and two-phonon states, and so on.  These states occupy a special
 place in nuclear physics. Theoreticians point out that many difficulties of
 the nuclear theory
 are concentrated  just on excited 0$^+$ states \cite{Sol89}.
 This conclusion was achieved already when spectra of only a few additional 0$^+$ states
 at low energies above the $\beta$ vibrational state in deformed nuclei were known:
the interacting boson model (IBM) \cite{IBM} and
the quasiparticle phonon model (QPM) \cite{QPM}
 have been used extensively for comparison with experimental data.
  Theories met difficulties already for consideration of the properties of
  the first excited states, for example, at explaining  the strong excitation of
  the first excited $0^+$ states in actinide nuclei.
  They represent the collective excitations different in character from the $\beta$ vibrations.
 Importance of the monopole and the quadrupole  pairing field was realized
 \cite{Grif71,Rij72,Bess72} while trying to explain  this observation.
 Garrett \cite{Gar01} reviewed the properties of the first excited 0$^+$ states
 in deformed nuclei and showed that only in a few nuclei
 the  states,   considered as
 $\beta$ vibrational,  met the original definition  \cite{Bohr98}.
 In all other nuclei they have a more complex structure.

The multiple $0^+$ states in deformed nuclei are so far  a new
challenge for all theoretical models. After the first observation  of multiple $0^+$ states in  $^{158}$Gd
\cite{Lesh02}, the  attempts were concentrated to understand the nature of these states in deformed
nuclei in general, and, in particular,   the  observed 13 excited $0^+$ states in $^{158}$Gd.
Simple calculations were performed by using the projected shell model \cite{Sun03}
and geometric collective model  \cite{Zam02}. The most popular approaches were applied
in the framework of the IBM \cite{Zam02}
and  the QPM \cite{Lo04}. Some other theoretical
approaches have been used to describe  0$^+$ states in different nuclei, for example,
within a model based on the Hamiltonian including the monopole pairing,
quadrupole-quadrupole, and spin-quadrupole interactions, all diagonalized in the random phase
approximation \cite{Mur10}. In fact, the nature of the $0^+$  states in  these
approaches are different.  For example, the role of octupole components in the formation of
$0^+$  states in actinides is radically different in the IBM and QPM \cite{Zam01,Zam03,Lo05}.
Nevertheless,  the models describe some properties of the energy spectra of  $0^+$ states and
the excitation cross sections.
In comprehensive theoretical efforts
to understand the nature of these $0^+$ excitations,  an extensive mapping of excited
$0^+$ states, information on the evolution of the abundance of $0^+$ states
in the entire region of deformed nuclei and on the dependence of the abundance of 0$^+$ states in the  excitation energy spectrum are important.

Excited 0$^+$ states are usually identified via  (p,t)  reactions  even in the complicate
and dense excitation spectra: they have a very  distinct angular distribution.
Early studies, for example in Ref.~\cite{Bal96}, were limited  relative
to the low excitation energies,
and a limited number of excited $0^+$ states were observed.
Intensive studies of the multiple $0^+$ states
were triggered by the observation of 12 excitations with  zero angular-momentum transfer via
the (p,t) reaction  in the odd nucleus $^{229}$Pa \cite{Lev94} and 13 excitations
in the even-even   nucleus $^{158}$Gd \cite{Lesh02}.  Then, many experiments were carried
out  through the (p,t) transfer campaign in the region of actinides
\cite{Wir04,Lev09,Lev13,Lev15,Spi13,Spi18} and
rare earths \cite{Mey06,Buc06,Bet09,Ili10,Ber13}.
A feature of some of these studies is that, simultaneously with $0^+$ states, many states
with other spins of both parities have also been identified. The total spectra
of $0^+$ states  and, presumably, the total spectra
of 2$^+$ and 4$^+$  states of a collective nature were  accumulated.

In deformed nuclei an excitation mode with angular momentum
$J^\pi$ splits into states distinguished by its projection $K$ quantum number, which
ranges from zero to $J$.  In addition to the states with structure $J^{\pi}K^{\pi}=0^+0^+$
 the states  with a more general  structure $J^{\pi}K^{\pi}=J^+0^+$ are expected.
States related to specific two-quasiparticle modes are expected above twice the pairing
gap energy. In addition, there are two-phonon excitation of collective modes,
the quadrupole, octupole and hexadecapole phonons.
The collective monopole pairing vibration  have also to be considered.
All these configurations will mix to some extent.
Thus, a significant number of 0$^+$ states with different structures should be observed in deformed nuclei.

\begin{figure}
\includegraphics[width=0.48\textwidth,clip]{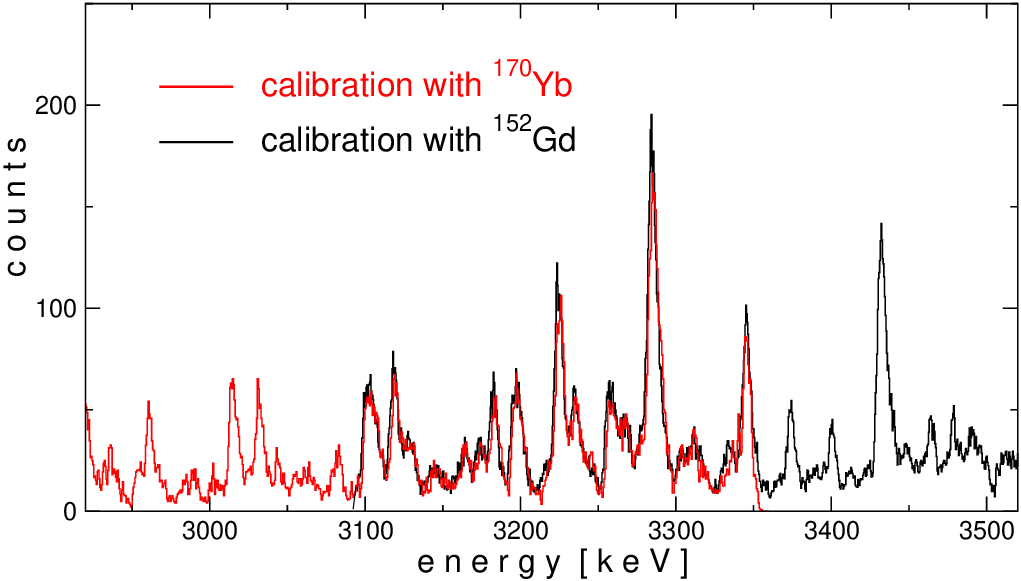}
\vspace{-0.0mm}
\caption{{\small The low- and high-energy spectra, calibrated
by levels of $^{170}$Yb and $^{152}$Gd, respectively.  Their matching
in the overlapping area demonstrate the accuracy of the calibration. }}
\label{Fig1_overlap}
\end{figure}

\begin{figure*}
\includegraphics[width=0.9\textwidth,clip]{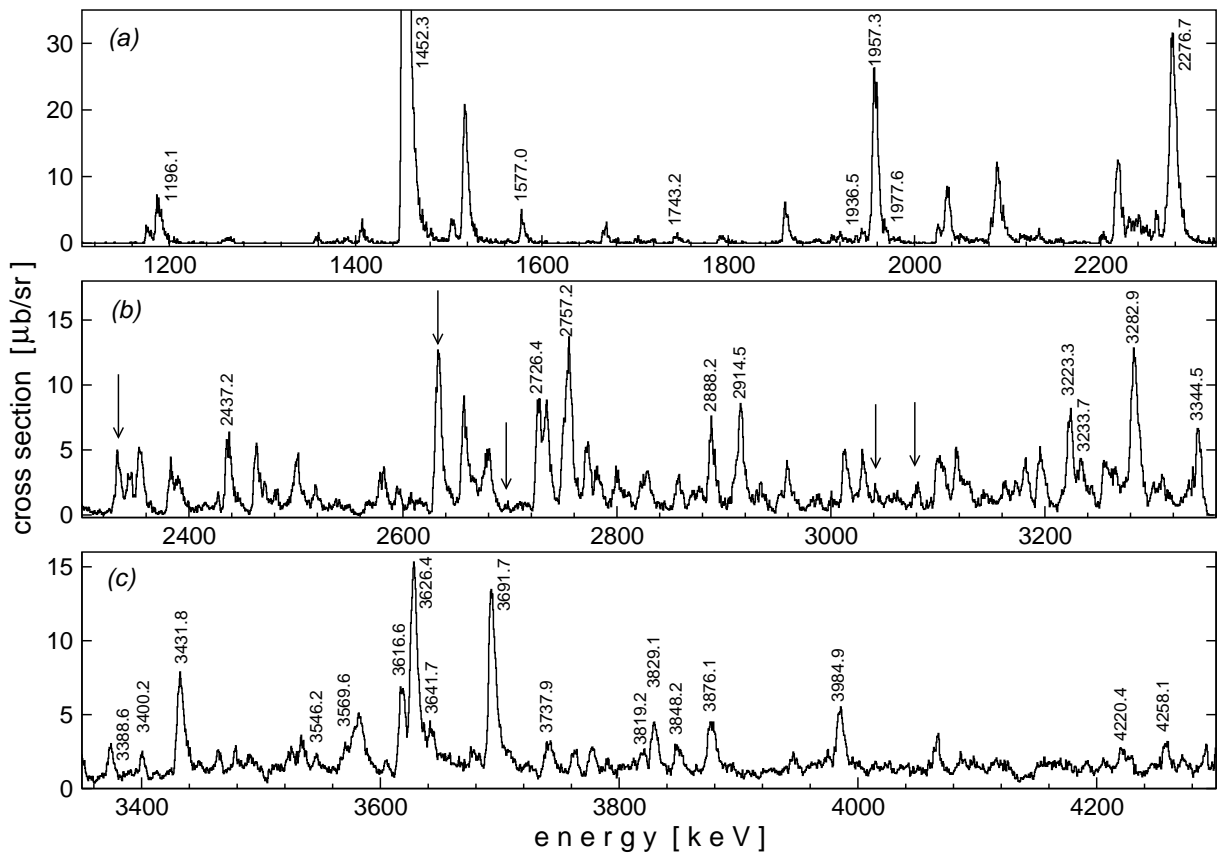}
\caption{{\small The triton spectrum from the $^{160}$Gd(p,t)$^{158}$Gd
reaction measured at angle 5$^\circ$.   The states assigned as 0$^+$ states
are labeled by their energies.  Arrows indicate the states for which
0$^+$ assignments in Ref.~\cite{Lesh02}  are not confirmed in this study. }}
\label{Fig2_Spec_1-4}
\end{figure*}

So far, almost all the studies of the 0$^+$ states have been performed for
an excitation energy below 3 MeV.  One attempt to expand
this range was undertaken for $^{230}$Th \cite{Lev09}.
In the region up to 4.5 MeV, the (p,t) spectra for two angles,
12.5$^\circ$ and 26$^\circ$ were measured. The ratio of cross sections
for 8 states was corresponding to the ratio for $0^+$ states.
However, this ratio  does not exclude the identification of $3^+$ and $6^+$ states.
The answer to the assumption
that the spectrum of $0^+$  states  is terminated in energy is still open.
The angular distributions up to approximately 4 MeV were also measured for $^{168}$Er \cite{Buc06}.
In the region from 3 to 4 MeV   a  steeply  rising cross section was
observed at small reaction angles  for 9 excitation energies. However, a sharp minimum
at about 17.5$^\circ$, which is also a distinguishing feature of $0^+$ excitations,
was absent in these angular distributions.
 Since such angular distributions
correspond also to 2$^+$ excitations in the DWBA  calculations
 with taking into account an indirect transfer in the (p,t) reaction \cite{Lev15},
 these 0$^+$ states can be assigned only tentatively.

Our initial aim was to carry out  the $^{160}$Gd(p,t)$^{158}$Gd  experiment
for observation  of the $0^+$ excitations in the region from 3 to 4.2 MeV, in
addition to the already observed  $0^+$ states below 3 MeV by  Lesher et al.
\cite{Lesh02}. However, some of the circumstances discussed below led to
the need to perform the experiment also for  lower energies.  Thus,
we identified 230 states with different spins. The purpose of this paper is
to present the results  for  $0^+$ states: we report  the existence
of 34  $0^+$ states  in one nucleus below the excitation energy of
 4.3 MeV. For four of them, including the 1952.4 keV state,
 the assignment is tentative. This number is the largest observed in any nucleus
and provides a unique opportunity for  testing new models on the nature of
$0^+$  excitations in nuclei. Results on new  $2^+$, $4^+$, and $6^+$
states will be presented in a forthcoming work.

\section{\label{sec:ExAnRe} Experiment details and   results}

The first experiment in the high energy region has been performed
at the Tandem accelerator of
the Maier-Leibnitz-Laboratory of the Ludwig-Maximilians-University
and Technical University of Munich  using a 22 MeV proton beam
on a 110 $\mu$g/cm$^2$ target of isotopically enriched $^{160}$Gd (98.10\%)
with a 14 $\mu$g/cm$^2$ carbon backing. Known impurities in the target material
consist of $^{158}$Gd (0.99\%), $^{156}$Gd (0.33\%), and $^{157}$Gd (0.44\%).
A 1.4 m long focal plane detector provides  the particle identification
of  the ejectiles of masses 1 - 4 in the high-precision Q3D spectrometer.
The resulting triton spectra have a resolution of 4 - 7 keV (FWHM) and
are background-free. The acceptance
of the spectrograph was 14.43  msr  for all angles, except for
the most forward angle 5$^\circ$, where it was
7.50  msr. Typical beam currents was about 1.0 $\mu$A.
The angular distributions of the cross sections were obtained from
the triton spectra at eight laboratory angles from 5$^\circ$ to 40$^\circ$.
The low energy spectra in the interval from 0 to 3.4 MeV
have been  also measured at angle 5$^\circ$  for three magnetic setting,
which are all overlapping with the neighboring regions.

 For the calibration of the energy scale, the triton spectra from
 the reactions  $^{154}$Gd(p,t)$^{152}$Gd  have been measured at
 the same magnetic  setting.  The high energy spectrum of $^{158}$Gd
 was calibrated with the known energies from $^{152}$Gd,
 while the lower energy part was calibrated, at first,
 using the energies of the 0$^+$ states assigned by Lesher et al. \cite{Lesh02}.
 When the high-energy spectrum was
shifted to the overlapping region  with the low-energy spectrum
 it occurs to be
impossible to combine the high and low energy
spectra.
The energy scales were different,
and a necessary shift
was found considerably different from the Q-value obtained
using the Atomic Mass Tables  \cite{AME16}.
To solve this problem, a second experiment was performed in
the low-energy region on the  125 $\mu$g/cm$^2$ target of $^{160}$Gd.
The acceptance of the spectrograph was 9.16 msr  for 6$^\circ$
and 15.94 msr for other angles.
The resulting triton spectra have a slightly lower resolution  of 8 - 9  keV (FWHM).
For the calibration of the energy scale, the triton spectra from
the reactions  $^{172}$Yb(p,t)$^{170}$Yb  are measured at
the same magnetic  settings.
The well-known levels  of $^{158}$Gd have been also used for calibration in
this energy interval.  As  seen from  Fig.~\ref{Fig1_overlap}, the spectra in
both energy intervals calibrated by the reactions $^{154}$Gd(p,t)$^{152}$Gd
and  $^{172}$Yb(p,t)$^{170}$Yb coincide perfectly in the overlapping  region,
which is an evidence for the accuracy of the calibration.
 Fig.~\ref{Fig2_Spec_1-4}(a-c) shows the triton spectrum  over the whole measured
  energy interval from 1.0 to 4.3 MeV, taken  at the detection angle of 5$^\circ$.
 Assigned 0$^+$ states are labeled by their energies in keV.
The states assigned as 0$^+$ states in Ref.~\cite{Lesh02} which are not confirmed in this study
are labeled by arrows.

\begin{table}
\caption \normalsize {Reaction Q-values obtained from the energy shifts of the $^{152}$Gd
and $^{170}$Yb spectra, relative to the $^{158}$Gd spectrum in the calibration procedure
 are compared with the Q-values calculated from the mass excesses. All data are given in keV.}\\
\label{Tab1_Qval}
\hspace{5mm}
\begin{ruledtabular}
\begin{tabular}{cccccc}\\
               & $^{152}$Gd    &       &   $^{158}$Gd &      &    $^{170}$Yb   \\
               \hline\\
      $\triangle E$        &          &1749.0 \it6&         & 1238.1 \it6 &        \\
      $Q$[$\triangle E$]     & 6671.5 \it9 && 4912.9 \it7     &&  6161.0 \it9 \\
      $Q$[AME]     & 6659.9 \it3  && 4912.9 \it7     &&   6152.3 \it6  \\
      $\triangle Q$      & 11.6 \it10   &&           &&      8.7 \it9\\
\end{tabular}
\end{ruledtabular}
\end{table}

In the course of analysis of the measured spectra we found that
the Q values for the (p,t) reactions on the $^{160}$Gd,
 $^{154}$Gd  and $^{172}$Yb targets are in disagreement with  the ones
calculated using the  Atomic Mass Tables \cite{AME16}.
They are given in Table~\ref{Tab1_Qval}.
The reaction Q-value for the $^{160}$Gd target  is used as the reference value.
This means that for this nucleus the Q value which was determined from the data
in the AME2016 was taken as a starting point in the calculations for other nuclei.
As seen from the table,  the Q value for the $^{154}$Gd  and $^{172}$Yb targets, which is
determined by the energy shift necessary to fit the peaks in the overlapping
region, differs from the value obtained by using the  atomic mass excesses.
Since the differences for  $^{154}$Gd  and $^{172}$Yb targets are positive and
close in values, the inaccuracy in the mass excess  refers most likely to $^{160}$Gd and/or $^{158}$Gd.

\begin{figure*}
\includegraphics[width=0.85\textwidth,clip]{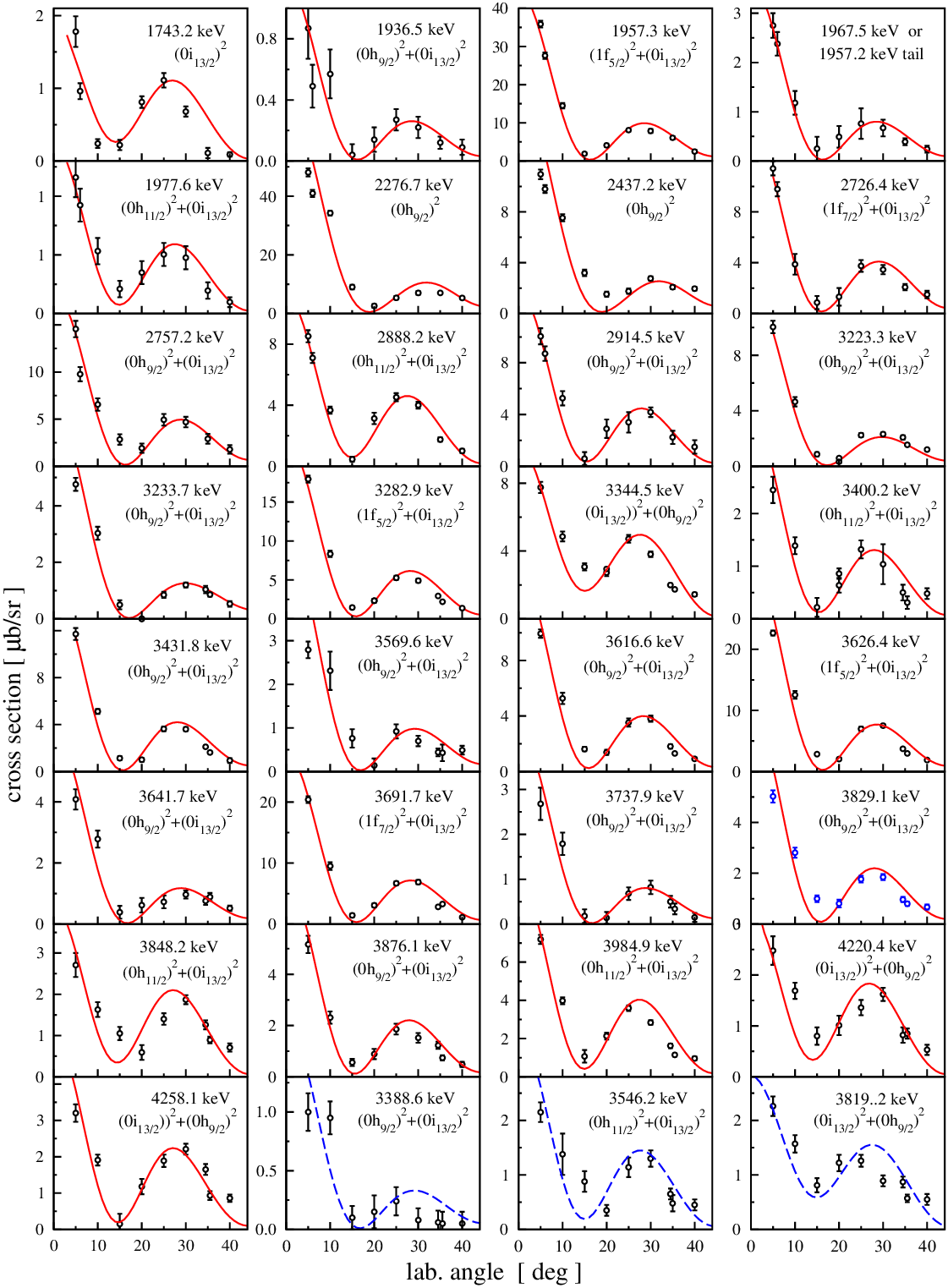}
\caption{{\small Angular distributions of assigned 0$^+$
states in $^{158}$Gd and their fit with the CHUCK3 one-step calculations.
The transfer configurations used in the calculations for
the best fit are shown for every state (see text for details).}}
\label{Fig3_results}
\end{figure*}

The analysis of triton spectra was performed  using the program GASPAN \cite{Rie91}.
The peaks in the spectra which are measured at 5$^\circ$ degree have been  identified for 230 levels,
though the peaks for  all eight angles were identified only for 162 levels.
 The resulting angular
distributions are shown for 0$^+$ states  in Fig.~\ref{Fig3_results}. Efficiency corrections
for the target thickness at different angles have been taken into account.

The observed angular distributions are compared with calculations using
the  distorted wave Born approximation (DWBA).  The coupled-channel approximation
(CHUCK3 code of Kunz \cite{Kun})  and the optical potential parameters suggested
by Becchetti and Greenlees \cite{Bec69} for protons and  by Flynn et al. \cite{Fly69}
for tritons have been used in the calculations.   Angular distributions of
the 0$^+$ states are reproduced very well by a one-step process, which simplifies the calculations.
The orbitals close to the Fermi surface have been used as the transfer
configurations.  For $^{158}$Gd and $^{160}$Gd  such configurations include the orbitals,
which correspond to those
 in the spherical potential, namely, $1h_{9/2}$, $2f_{5/2}$,
$1i_{13/2}$, and $1h_{11/2}$.
The DWBA angular distributions depend to some extent on the transferred configurations.
The most noticeable difference is obtained for the angular distribution at
the $(1i_{13/2})^2$  transfer configuration.
For  other configurations the difference consists in a different height of
the maximum at about 20$^\circ$ and minor displacements of minimum.
In addition, since the excited $0^+$ state must consist of many terms in the wave function with
a coherent summing of individual amplitudes, this difference allows  to obtain
a better fit to the experimental angular distributions  using mixed configurations.
They are shown in Fig.~\ref{Fig3_results}. Only two transfer components are shown:
the first one is the main constituent, while the second one improves the fit to the peak at 20$^\circ$ and to the minimum at about 15 - 18$^\circ$, its admixture does not exceed 10\%.
\newcolumntype{d}{D{.}{.}{3}}
\begin{table*}[]
\caption \normalsize {Results of the present (p,t) experiment are compared with previous studies.
The first column shows the energies measured by different methods and compiled in Ref.~\cite{Nic17}.
Next three columns show  energies, relative (p,t) cross sections at 6$^o$, and spins
from  Ref.~\cite{Lesh02}. The last three columns show  the present results: energies,
absolute (p,t) cross sections at 5$^o$, and spin assignments.  The errors of the differential cross sections are statistical, and an additional error of 10\% should be taken into
account due to the uncertainty in the thickness of the targets used.}\\
\label{Tab2_results}
\hspace{5mm}
\begin{ruledtabular}
\begin{tabular}{l lc c lcc}\\
{NDS Ref.~\cite{Nic17}}&\multicolumn{3}{c}{Results of Lesher et al. Ref.~\cite{Lesh02}}&\multicolumn{3}{c}{Results of present study}\\
\smallskip\\
{$E_{exp}$}(keV)& {$E_{exp}$}(keV)&{$d\sigma/d\Omega$}(rel)& $I^\pi$ &{$E_{exp}$}(keV)&{$d\sigma/d\Omega$}($\mu b$) & $I^\pi$\\

\smallskip\\
\hline\\
0.0          &  0.0 \it6    & 1000 \it8  & 0$^+$ &0.0 \it3 & 1320 \it12 & $0^+$\\
1196.165 \it8 & 1194.8 \it13 & 3.7 \it 6 & 0$^+$ &1196.1 \it8 & 3.1 \it4&$0^+$\\
1452.352 \it6 & 1452.4 \it 6 & 305 \it 6 & 0$^+$ &1452.3 \it3 & 389 \it6 &$0^+$\\
1576.930 \it16 & 1577.0 \it12 & 5.4 \it7 & 0$^+$ &1577.0 \it4 & 5.6 \it6 &\\
1743.145 \it14 & 1742.7 \it9 & 0.6 \it3  & 0$^+$ &1743.2 \it5 & 1.8 \it3&$0^+$\\
1935.5 \it6 &&&&1936.5 \it15& 0.9 \it3    &$0^+$\\
1952.424 \it 25 &1953.5 \it6& 30.8 \it14 & 0$^+$ &1952.4$^*$ & 0.4 \it3&\\
1957.424 \it25 & 1960.1 \it38 & 3.2 \it5 & 0$^+$ &1957.3 \it3 & 35.9 \it 9&$0^+$\\
1972 \it3 & 1972.2 \it31 & 0.4 \it2      & 0$^+$ &1977.6 \it 8 & 1.2 \it 2 & $0^+$ \\
2276.02 \it3 & 2277.3 \it22 & 39.6 \it22 & 0$^+$ &2276.7 \it4 & 48.2 \it 12 &$0^+$ \\
             &2338.0 \it8 & 10.7 \it7    & 0$^+$ &2333.4 \it5 & 6.7 \it4 & $4^+$ \\
             &&            &              &2437.2 \it 4 & 11.0 \it 4 &$0^+$ \\
             &2643.4 \it8 & 18.1 \it10   & 0$^+$ &2632.7 \it4 & 20.0 \it8 & $4^+$\\
                            &&               &         &2643.1 \it 5 & 2.3 \it 3 &$2^+$ \\
              & 2688.8 \it8 & 1.7 \it10   & 0$^+$ &2695.5 \it8 & 0.8 \it3& $2^+$\\
                &&               &         &2726.4 \it 4 & 11.4 \it 6 &$0^+$ \\
                &&&&2757.2 \it 4 & 14.6 \it 9 &$0^+$ \\
             &2911.2 \it 11 & 8.7 \it13 & 0$^+$ &2888.2 \it4 & 8.5 \it5 & $0^+$\\
2913.4 \it7               & &&&2914.5 \it 5 & 10.0 \it 6 &$0^+$ \\
             &3076.7 \it16 & 2.9 \it49 & 0$^+$ &3041.7 \it8 & 1.5 \it3 & ($2^+$) \\
3080.0 \it6             & 3109.9 \it11 & 1.2 \it5 &0$^+$ & 3079.2 \it5 & 2.1 \it3 & ($?$)\\
            & &&&3223.3 \it3 & 10.0 \it5 & $0^+$ \\
3234.5 \it5          &   &&&3233.7 \it4 & 4.8 \it3 & $0^+$ \\
            & &&&3282.9 \it5 & 18.0 \it6 & $0^+$ \\
             &&&&3344.5 \it5 & 7.7 \it 4 & ($0^+$) \\
             &&&&3388.6 \it9 & 1.0 \it2 & ($0^+$) \\
             &&&&3400.2 \it9 & 2.5 \it3 & $0^+$ \\
             &&&&3431.8 \it8 & 10.3 \it5 & $0^+$ \\
             &&&&3546.2 \it7 & 2.0 \it3 & $0^+$ \\
3570.7 \it6   &          &&&3569.6 \it7 & 2.8 \it3 & $0^+$ \\
             &&&&3616.6 \it8 & 9.9 \it5 & $0^+$ \\
3626.9 \it6   &          &&&3626.4 \it8 & 22.7 \it6 & $0^+$ \\
             &&&&3641.7 \it8 & 4.1 \it4 & $0^+$ \\
             &&&&3691.7 \it8 & 20.4 \it6 & $0^+$ \\
             &&&&3737.9 \it11 & 2.7 \it6 & $0^+$ \\
3819.8 \it     &        &&&3819.2 \it7 & 2.2 \it3 & ($0^+$) \\
             &&&&3829.1 \it6 & 5.0 \it4 & $0^+$ \\
             &&&&3848.2 \it8 & 2.6 \it3 & $0^+$ \\
             &&&&3876.1 \it6 & 5.2 \it4 & $0^+$ \\
             &&&&3984.9 \it6 & 7.2 \it4 & $0^+$ \\
             &&&&4220.4 \it6 & 2.5 \it4 & $0^+$ \\
             &&&&4258.1 \it6 & 3.3 \it4 & $0^+$ \\
\end{tabular}
\end{ruledtabular}
\vspace{5mm}
$^*$ The peak at 1952.4 keV is hidden by much more strong peak at 1957.3 keV.
Its strength was estimated by fixing energy of 1952.4 keV in the process of fitting.
\end{table*}

 In Fig.~\ref{Fig3_results}, the experimental data   are given in $\mu$b/sr
and  their values
are plotted with the error bars while the Q-corrected CHUCK3 calculations
are shown with full lines. The solid (red) lines present the firm assignments
and the dashed (blue) lines show tentative assignments.
The results of this study concerning 0$^+$ states as compared
with previous studies  are  collected in Table \ref{Tab2_results}.
The  high-precision study  of the (p,t) reaction on $^{160}$Gd was performed
by Lesher et al. \cite{Lesh02}.
They confirmed the definite assignment of three excited 0$^+$ states
at 1196.1, 1452.3, and 1743.1 keV
as well as four tentative 0$^+$ assignments for 1952.4, 1957.4, 1972.2, and  2688.8 keV
from the (n,$\gamma$) reaction \cite{Gre78}.
Additionally, they found  seven new 0$^+$  states
at 1577, 2277, 2338, 2643, 2911, 3077, and 3110 keV.
The 0$^+$ assignment at 1577 and 2277 keV was strengthen by the analysis
of gamma rays from the (n,$\gamma$)  reaction \cite{Gre78}.
However,  later  in the next paper by Lesher et al. \cite{Lesh07}
for study of 0$^+$ states in the (n,n$^{\prime}\gamma$) reaction,
no $\gamma$-rays were detected as decaying the level 1577 keV.
Therefore, the corresponding peak in the observed triton spectrum is interpreted
as an existing excitation through the $^{156}$Gd(p,t)$^{154}$Gd reaction
on the $^{156}$Gd impurity  in the target
\cite{Lesh07}. The observed cross section 3.8 $\mu$b/sr is only slightly larger
than the calculated 2.7 $\mu$b/sr under the assumption of identical cross sections
for the ground state excitations in $^{158}$Gd and $^{154}$Gd nuclei.

\begin{figure}
\includegraphics[width=0.45\textwidth,clip]{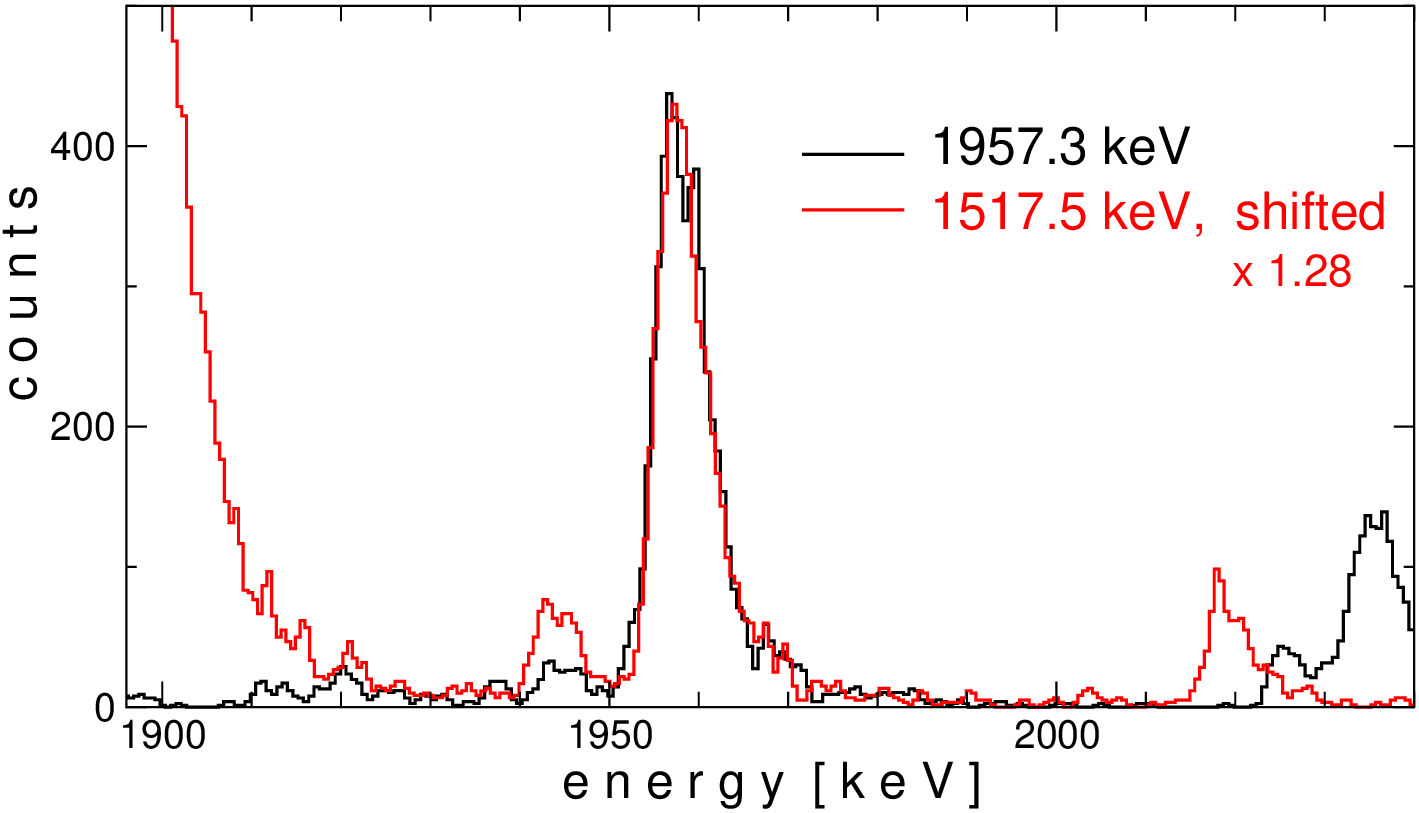}
\vspace{-0.0mm}
\caption{{\small The  shape of peak at 1957.3 keV  and the peak at
1517.5 keV shifted and normalized to overlap with the 1957.3 keV peak.}}
\label{Fig4_1960}
\end{figure}

For the states below 1743.2 keV, only the absolute cross section
is shown in Table~\ref{Tab2_results}
as the result of our analysis, since their angular distributions
were not measured.  The angular distributions confirm the 0$^+$ assignment
for the state 1743.2 keV and,  for the first time, for the 1936.5 keV state,
although with a very small cross section in the latter case.
The strong 1953.5 keV  peak   was attributed
by Lesher et al. \cite{Lesh02} to the excitation of the 1952.4 keV state
and the 1960.1 keV peak
to the state  1957.4 keV  known from  the previous publications, e.g. Ref.~\cite{Gre78}.
According to our study with correct calibration, the strong peak
is observed at the 1957.3 keV and
it should be attributed to the excitation of the known state 1957.4 keV.
 Lesher et al. in another publication \cite{Lesh07} using an analysis of
the $\gamma$-rays from the (n,n$^{\prime}\gamma$) reaction
confirm their private communication with Bucurescu and Meyer and concluded that
the putative 1953.5 keV level actually has an energy of 1957.0 keV.
Moreover,  Bucurescu and Meyer find no evidence of the suggested 1960.1 keV level.
In fact, when the energy of a strong peak is shifted in Ref.~\cite{Lesh07},
the same shift should be applied also for this weak peak,
giving an energy of about 1964 keV.
Our calibration procedure produces an energy of 1966.5 keV for this peak.
 Nevertheless an additional test
confirms the  conclusion in Ref.~\cite{Lesh07}. Fig.~\ref{Fig4_1960} shows
an overlapping of the 1957.3 keV peak with
another single peak from the spectrum, such that their tails practically coincide.
  Thus, the small peak  distinguished at 1960.1 keV
 by Lesher et al. (1966.5 keV in our case) is most likely
a result of the tail from the 1957.3 keV strong peak.
The same angular distribution for the peak  at 1957.3 keV and for
the  doubtful  peak at 1966.5 keV (see Fig.~\ref{Fig3_results})
 is an additional argument.

A 0$^+$ level at 1952.34$\pm$0.05 keV had been tentatively
proposed from the neutron capture data \cite{Gre78}.
A confirmation of this would be the observation of the 0$^+$
state in the (p,t) reaction. However,
the excitation of the  1952.4 keV state is very weak and,
for that,  measurements
of the angular distribution, as it turned out, are not possible.
Therefore, our data cannot confirm a 0$^+$ assignment for this state
and only a tentative spin can be inferred from the gamma ray data.
The 0$^+$ assignment  is not supported for the 1972.2 keV state  in Ref.~\cite{Lesh02}.
Instead we found a weak and spread peak with the energy  determined
as 1977.6  keV, and its angular distribution supports the 0$^+$ assignment.

\begin{figure}
\includegraphics[width=0.50\textwidth,clip]{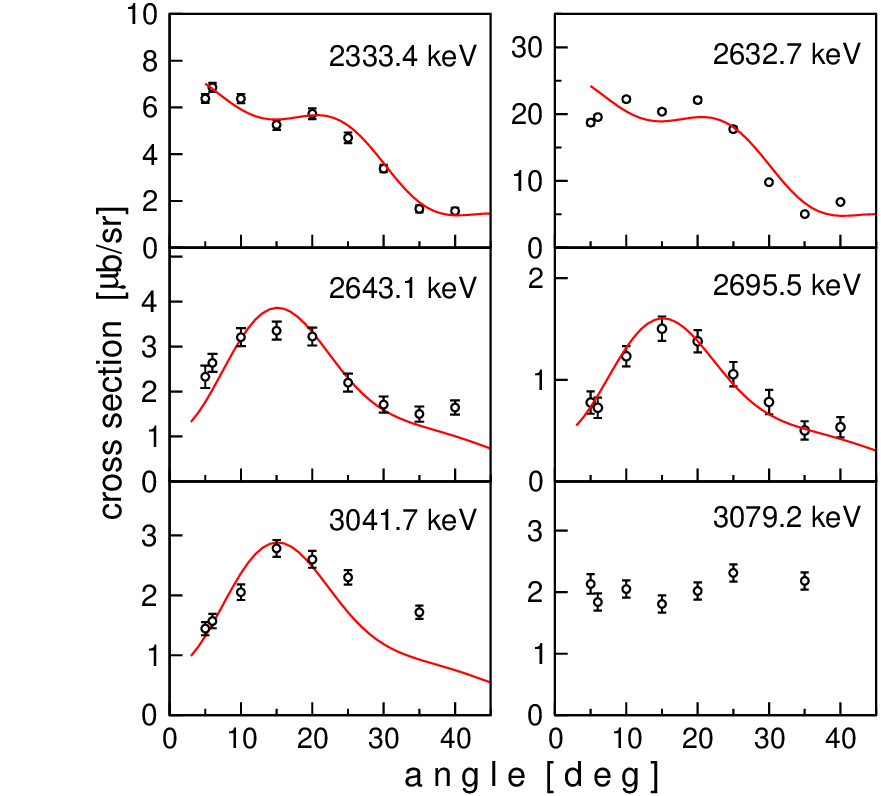}
\vspace{-0.0mm}
\caption{{\small The angular distributions for the states assigned
as 0$^+$ excitations in Ref.~\cite{Lesh02}. Our results do not confirm
these assignments. }}
\label{Fig5_contra}
\end{figure}

In the part of the higher energy spectrum
we support the 0$^+$ assignment for two states out of the seven 0$^+$ levels,
all seven assigned in Ref.~\cite{Lesh02}.  The  energies of all these states
differ in our study from those given in Ref.~\cite{Lesh02}.
The reason resides in the different calibrations we used,
whose details are given above.
The details  of calibration used by Lesher et al. \cite{Lesh02} are not available
in their publication.  Therefore, in the following discussion, we will refer
to the energies determined in our study.
Using the present calibration, we support  the 0$^+$ assignment
for the states 2276.4 and 2888.2 keV.
The angular distributions obtained in our study for the rest of the five states
reported as 0$^+$ states in Ref.~\cite{Lesh02} are presented in Fig.~\ref{Fig5_contra}
 together with  DWBA calculations.
They allowed assignments of spin
4$^+$ for the states 2333.4 and 2632.7 keV,  2$^+$ for the state  2695.5 and 3041.7 keV.
The angular distribution for
the state 3079.2 keV does not allow the definite assignment.
At the same time we found four new 0$^+$ states at 2437.2, 2726.4, 2757.2, and 2914.5 keV.
As  seen from  Fig.~\ref{Fig3_results}, their angular distributions indicate clear 0$^+$ assignments.


\begin{figure}
\includegraphics[width=0.45\textwidth,clip]{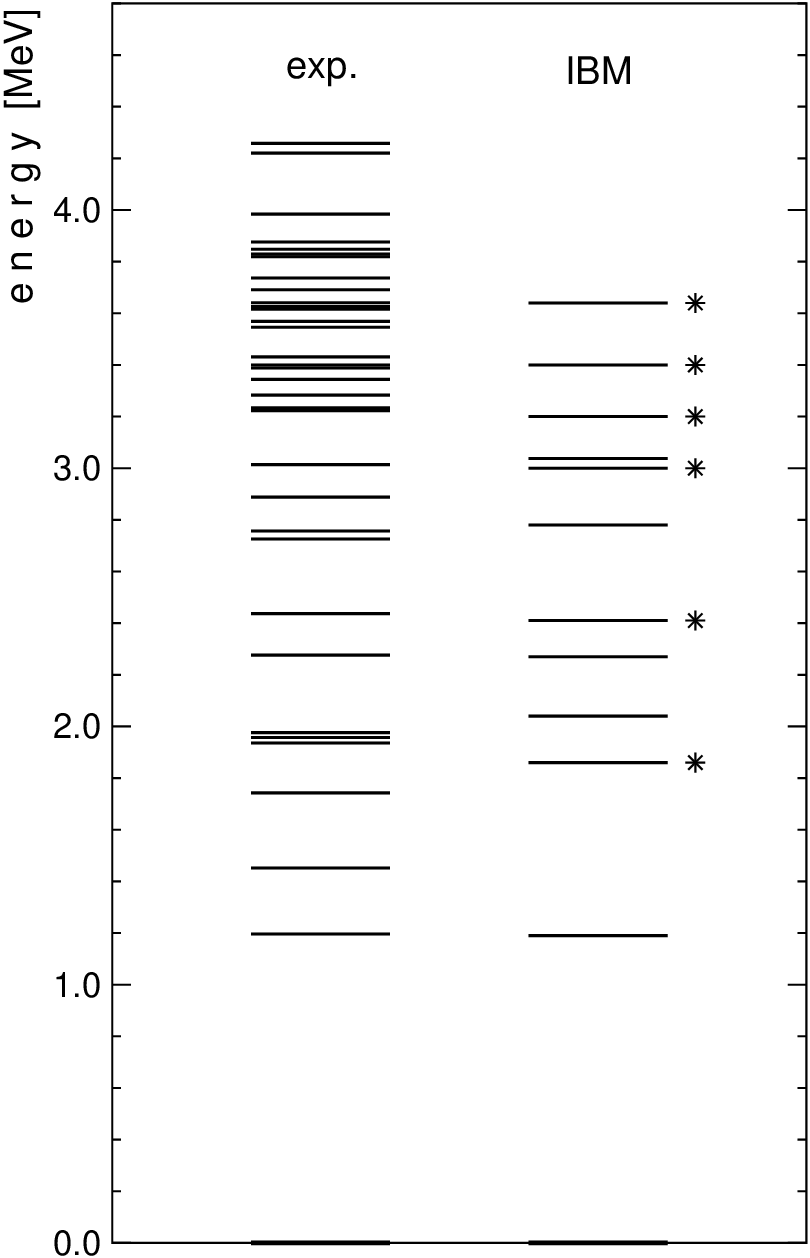}
\vspace{-0.0mm}
\caption{{\small  The experimental 0$^+$ excitation energies compared with
the calculations within the spdf-IBM. Stars on the calculated 0$^+$ states
indicate levels with doubly octupole character.}}
\label{Fig6_0plus_IBM}
\end{figure}

\begin{figure}
\includegraphics[width=0.48\textwidth,clip]{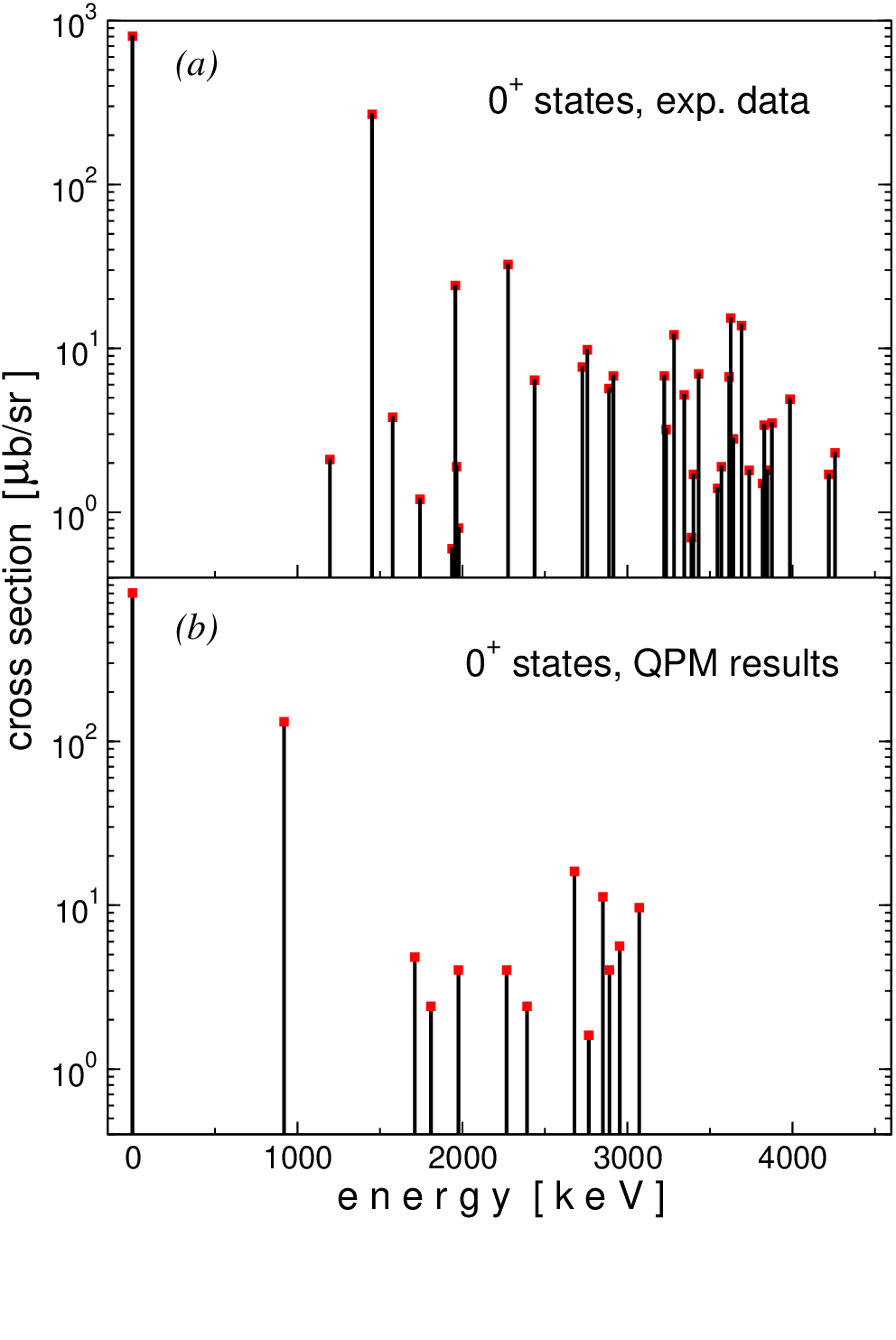}
\vspace{-10.0mm}
\caption{{\small  The (p,t ) cross sections at the angle 5$^\circ$
for 0$^+$ states in $^{158}$Gd: experimental data (a)
and calculated in the framework of the QPM (b). }}
\label{Fig7_0plus}
\end{figure}

There are, however, disagreements between some of  our data
and the results by Lesher et al. \cite{Lesh07} from study of the 0$^+$ states
in the (n,n$^{\prime}\gamma$) reaction.
The aim of this study was to define  the collective
nature of 0$^+$ excitations assigned in their previous work using the (p,t) reaction.
The main way of decay of the low-lying 0$^+$ states is to the first excited
state 2$^+$ at energy of 79.5 keV.  Therefore, $\gamma$ rays of the corresponding
energies were found in the $\gamma$ spectrum and their properties were investigated.
This study confirms the data for
the 1743.2, 1957.3 (with the energy correction discussed above) and 2276.7 keV.
Study of the coincidences of the $\gamma$ rays feeding and de-exciting 0$^+$ states
confirms these assignments.

The disagreements start from the state with an energy of 2333.4 keV,
as determined in the present study, and with 2338.0 keV in Ref.~\cite{Lesh02}.
From the observation of the
de-exciting $\gamma$ rays Lesher et al. have identified a 0$^+$ state at energy 2340.0 keV
although  the gamma ray decaying to the first 2$^+$ state was not observed.
 They concluded that the
 state of 2340.0 keV and that of 2338.0 keV observed in the (p,t) reaction
 is the same state.
However, the energy of corresponding peak with our calibration is 2333.4 keV and
as seen from Fig.~\ref{Fig5_contra} the angular distribution for this level corresponds
to the 4$^+$ spin assignment.  The energy  2338.0 keV and, especially, 2340.0 keV
are excluded additionally by the triplet of peaks in the triton spectrum
with  the two known  energies of 2355.0, 2344.7 keV \cite{Nic17} and with
a third energy level in question.
The triplet looks almost equidistant, which allows one to obtain the energy of the peak
in question from the interval between the peaks of 2355.0 and 2344.7 keV.
The obtained energy is 2334.4 keV that is close to the value found in this study.
It can be assumed that the energy of 2340.0 keV  refers to the 2$^+$ state
known from the (d,p) reaction \cite{Nic17}, but not to the 0$^+$ state.

The 0$^+$ state at energy 2644.2 keV  was identified  using the observation
of the 2564.7 keV $\gamma$ ray and the excitation function.
This 0$^+$ state was assigned in the (p,t) reaction in Ref.~\cite{Lesh02}.
The 2564.7 keV $\gamma$ ray
was attributed to the transition from the 0$^+$ state to the 2$^+_1$ state
at the 79.4 keV energy \cite{Lesh07}. However, the angular distribution of
the $\gamma$ ray of 2564.7 keV is not completely isotropic
which already excludes a definite  0$^+$ assignment.
Finally, the real energy of the corresponding
peak as follows from our calibration is 2632.7 keV and
as seen from Fig.~\ref{Fig5_contra} the angular distribution for this level corresponds
to the 4$^+$ spin assignment. Since the coincidences were not measured,
it is not obvious that the $\gamma$ rays of 2564.7 keV refer to the de-excitation
of the 0$^+$ state into the 2$^+_1$ state.  Therefore, the situation
with the identification of this state is not clear.
 Perhaps, the 2564.7 keV $\gamma$ ray refers to the de-excitation of
the 2643.1 keV level, seen in the present study with close energy
to the putative 2644.2 keV state,
which is identified, however,  as the 2$^+$ level (see. Fig.~\ref{Fig5_contra}).

In a similar way, the 2832.0 keV  $\gamma$ transition was used to identify
the 0$^+$ state at an energy of 2911.5 keV, also assigned as 0$^+$ state
in the (p,t) reaction.
However, again the real energy of the corresponding peak is 2888.2 keV.

In addition,  we found 20 new 0$^+$ states in the energy interval
from 3200 to 4300 keV.  This energy region was not investigated so far in
the (p,t) reaction.  The total number of 0$^+$ states detected
in one nucleus equals now 34, which is
the largest of such states observed so far.
For three of them the  0$^+$ assignments are tentative.
For some of these states,
their energies  observed in the (n,$\gamma$) reaction were found to
be close within the error limits.  Apart from the energies, there is
no other information about these states. Therefore, one can not be sure that these
states and the ones observed in the (p,t) reaction are the same,
 although the close proximity of the energies obtained in the two
 independent experiments are supporting the validity of our calibration.

 As already mentioned,  theoretical models have relatively modest results
 describing the spectra of multiple 0$^+$ excitations. No attempt was made
to fit individual 0$^+$ states and, therefore, no predictions
of the 0$^+$ states having a correspondence with
 the specific experimental states.  The point of the calculations
was rather to see a number of 0$^+$ excitations in the energy range up to
about 3 MeV, and a general trend in the  cumulative  cross  section
with increasing energy.  Such calculations were performed both within
the framework of the QPM  and the  spdf-IBM,  in particular,
for $^{158}$Gd \cite{Zam02,Lo04}.  The IBM calculations yields
a number of  0$^+$ states close to the experimental ones below 3 MeV,
and many of the 0$^+$ states of two-phonon octupole character,
as shown in Fig.~\ref{Fig6_0plus_IBM}.
The spdf-IBM failed to reproduce the increasing density
of 0$^+$ states above 3 MeV.
In addition, several other 0$^+$ states at higher excitation energy are
calculated in Ref.~\cite{Zam02}, amounting to 23 excited 0$^+$ states
below 4 MeV. Therefore, spdf-IBM reproduces at least partially.
The cross sections were not calculated in this publication
since only the use of an extended Hamiltonian
allows to perform such calculations \cite{Pas10}.

The cross sections were calculated in the framework of the QPM.
The experimental spectra of 0$^+$ states, as compared to the calculated ones,
are shown in Fig.~\ref{Fig7_0plus}. The QPM  predicts a  number of
0$^+$ states which are close to the one observed below 3 MeV.
However, this model  fails in the cross section calculation
for the first excited state. This state is excited very
weakly, that may indicate  its $\beta$-vibrational nature.
A large cross section (33\% of the cross section for g.s.)
is observed for the second excited 0$^+$ state,
which is evidence of the similarity of its structure
to the structure of the ground state. In contrast to this, the QPM predicts
strong excitation just for the first excited 0$^+$ state, that shows
its resemblance  with the ground state, and very weak
excitations for all other 0$^+$ states.
Six of the QPM 0$^+$ states  (mostly the lowest) have a one-phonon character.
Other states at higher excitation energy contain large, and, in many cases, the dominant two-phonon components.
 They are built on the collective octupole phonons  almost in all cases,
 in agreement  with the IBM calculation \cite{Zam02,Lo04}.

New experimental data in the extended energy region represent an excellent opportunity to test these and other nuclear models.
There is one additional aspect of such studies.
The QPM  predicts an increase of the number of  0$^+$ states and a decrease
of their excitation cross sections in the (p,t) reaction
with increasing  the excitation energy \cite{Lo05}.
Their structure becomes more  complicated   and octupole components in the wave
function play an increasing role.  The  experimental spectrum of 0$^+$ states presented
in Fig.~\ref{Fig7_0plus}
demonstrates  a somewhat different picture.  A bump of states is observed in the region
between 3.2 and 4.0 MeV and, if there is no termination of spectrum,
 a  drop in the magnitude of the density of 0$^+$ states is then seen
 in experimental data.

\section{Conclusion}

We carried out a new high-precision (p,t)
reaction on an isotopically enriched target of $^{160}$Gd  which
 allowed the identification of  30 excited 0$^+$ states below 4.3 MeV
 in the spectrum of $^{158}$Gd.  Thus, the total number of 0$^+$ states
 in this nucleus is increased now up to 34.
Such abundance of 0$^+$ states have not previously been observed
in  any nucleus investigated so far.
The $^{158}$Gd was the nucleus for which information on the multiple 0$^+$ excitation
was published for the first time in Ref.~\cite{Lesh02}.
The new information can be interesting, especially among theoreticians,
because several models were applied in an attempt
to understand the nature of these states.  Much richer new
information will be of no less interest  for  theoreticians since
the observation of thirty four 0$^+$ states in one nucleus
is the strongest challenge  to our understanding of these excitations.
In a forthcoming  analysis of the obtained experimental data  the 2$^+$ and 4$^+$ states
 and possible other  states of the negative parity will be assigned.
 As in our previous
publications this can allow to build collective bands with
the 0$^+$ states  as  band-heads  which will bring further support
for the collectivity of these states.
The data from the (p,t) reaction are interesting in one more aspect.
As noted above, complete or almost complete sequences of states
of the collective nature with a definite  $J^\pi$ are available from this reaction.
This allows  to carry out a statistical analysis of these spectra
with the aim of clarifying the measure of order and chaos in
collective spectra \cite{Lev18,Mag18}. Moreover, such studies are helpful
in the formation of sequences of states that can be interpreted
as collective bands based on 0$^+$ and other states.
Collective bands with different $K$ for the 2$^+$ and 4$^+$ band heads can be formed,
and this opens a new possibility  to investigate the $K$-symmetry breaking \cite{Mag18}.

\section{Acknowledgments}

We are grateful to D. Bucurescu for the useful discussions and for providing
the helpful analyzed data from the second (p,t) experiment they have performed.
The work was supported in part by the Romanian project PN 18090102F2.
We thank also the operators at MLL for excellent beam conditions.



\begin{thebibliography}{99}

\bibitem{Sol89}   V.~G.~Soloviev, Theory of Atomic Nuclei:
            Quasiparticles and Phonons (Inst.Phys., Bristol, Philadelphia, 1992).

\bibitem{IBM} R.~F.~Casten and D.~D.~Warner, Rev. Mod. Phys.
             {\bf 60} 389 (1988) and references therein.

\bibitem{QPM} V.~G.~Soloviev, A.~V.~Sushkov, and N.~Yu.~Shirikova,
              Nucl. Phys. {\bf 568}, 244 (1994);
J. Phys. G: Nucl. Part. Phys. {\bf 20}, 113 (1994); Phys. Rev. C {\bf 51},
551 (1995);
Phys. At. Nucl. 59, 51 (1996); Phys. At. Nucl. 60, 1599 (1997).

\bibitem{Grif71} R.~E.~Griffin, A.~D.~Jackson, and A.~B.~Volkov,
            Phys.Lett. {\bf 36B}, 281 (1971).

\bibitem{Rij72} W.~I.~van~Rij and S.~H.~Kahana, Phys.Rev.Lett. {\bf 28}, 50 (1972).

\bibitem{Bess72} D.~R.~Bess, R.~A.~Broglia, and B.~Nilsson,
                 Phys. Lett. B{\bf 40}, 338 (1972).

\bibitem{Gar01}  P.~E.~Garrett, J. Phys. G: Nucl. Part. Phys. {\bf 27} (2001) R1 – R22.

\bibitem{Bohr98}   A.~Bohr and B.~R.~Mottelson, Nuclear Structure
                 (World Scientific, Singapore, 1998), Vol. 2.

\bibitem{Lesh02} S.~R.~Lesher, A.~Aprahamian, L.~Trache, A.~Oros-Peusquens, S.~Deyliz,
                A.~Gollwitzer, R.~Hertenberger, B.~D.~Valnion, G.~Graw,
                Phys.Rev. C {\bf 66}, 051305 (2002).

\bibitem{Sun03} Y.~Sun, A.~Aprahamian, J.-Y.~Zhang, C.-T.~Lee,
Phys.Rev. C {\bf 68}, 061301 (2003).

\bibitem{Zam02} N.~V.~Zamfir, J.-Y.~Zhang, R.~F.~Casten, Phys.Rev.
C {\bf 66}, 057303 (2002).

\bibitem{Lo04}  N.~Lo~Iudice, A.~V.~Sushkov, N.~Yu.~Shirikova,  Phys.Rev.
C {\bf 70}, 064316 (2004).

\bibitem{Mur10}   Murat Gerceklioglu, Phys. Rev. C {\bf 82}, 024306 (2010).

\bibitem{Zam01}  N.~V.~Zamfir and D.~Kusnezov,
Phys. Rev. C {\bf 63}, 054306 (2001).

\bibitem{Zam03} N.~V.~Zamfir and D.~Kusnezov,
Phys. Rev. C {\bf 67}, 014305 (2003).

\bibitem{Lo05} N.~Lo~Iudice, A.~V.~Sushkov, and N.~Yu.~Shirikova,
              Phys. Rev. C {\bf 72}, 034303 (2005).

\bibitem{Bal96}  H.~Baltzer, J.~de Boer, A.~Gollwitzer, G.~Graw, C.~G\"unther,
                 A.~I.~Levon, M.~Loewe, H.~J.~Maier, J.~Manns, U.~M\"uller,
                 B.~D.~Valnion, T.~Weber, and M.~W\"urkner, Z. Phys. A{\bf 356}, 13 (1996).

\bibitem{Lev94}   A.~I.~Levon,  J.~de Boer, G.~Graw, R.~Hertenberger,
    D.~Hofer, J.~Kvasil, A.~L\"osch, E.~M\"uller-Zanotti, M.~W\"urkner, H.~Baltzer,
    V.~Grafen, and C.~G\"unther,  Nucl. Phys. A{\bf 576}, 267 (1994).

\bibitem{Wir04} H.-F.~Wirth, G.~Graw, S.~Christen, D.~Cutoiu, Y.~Eisermann,
    C.~G\"unther, R.~Hertenberger, J.~Jolie, A.~I.~Levon, O.~M\"oller, G.~Thiamova, P.~Thirolf,
    D.~Tonev, and N.~V.~Zamfir, Phys. Rev. C {\bf 69}, 044310 (2004).

\bibitem{Lev09} A.~I.~Levon, G.~Graw, Y.~Eisermann, R.~Hertenberger, J.~Jolie,
    N.~Yu.~Shirikova, A.~E.~Stuchbery, A.~V.~Sushkov, P.~G.~Thirolf, H.-F.~Wirth, and N.~V.~Zamfir,
    Phys. Rev. C {\bf 79}, 014318 (2009).

\bibitem{Lev13} A.~I.~Levon, G.~Graw, R.~Hertenberger, S.~Pascu, P.~G.~Thirolf,
    H.-F.~Wirth, and P.~Alexa,  Phys. Rev. C {\bf 88}, 014310 (2013).

\bibitem{Lev15} A.~I.~Levon, P.~Alexa, G.~Graw, R.~Hertenberger, S.~Pascu, P.~G.~Thirolf, H.-F.~Wirth, Phys.Rev. C {\bf 92}, 064319 (2015).

\bibitem{Spi13}  M.~Spieker, D.~Bucurescu, J.~Endres, T.~Faestermann, R.~Hertenberger, S.~Pascu,
                 S.~Skalacki,  S.~Weber, H.-F.~Wirth, N.V.~Zamfir, and A.~Zilges,
                 Phys. Rev. C {\bf 88}, 041303(R) (2013).

\bibitem{Spi18} M.~Spieker, S.~Pascu, D.~Bucurescu, T.~M.~Shneidman, T.~Faestermann,
                R.~Hertenberger, H.-F.~Wirth, N.~V.~Zamfir, and A.~Zilges, Phys. Rev.
                C {\bf 97}, 064319 (2018).

\bibitem{Mey06}  D.~A.~Meyer, V.~Wood, R.~F.~Casten, C.~R.~Fitzpatrick, G.~Graw, D.~Bucurescu,
                 J.~Jolie,  P.~von~Brentano, R.~Hertenberger, H.-F.~Wirth, N.~Braun, T.~Faestermann,
                 S.~Heinze, J.~L.~Jerke,  R.~Kr\"{u}cken, M.~Mahgoub, O.~M\"{o}ller,
                 D.~M\"{u}cher, and C.~Scholl, Phys. Rev. C {\bf 74}, 044309 (2006).

\bibitem{Buc06}  D.~Bucurescu, G.~Graw, R.~Hertenberger, H.-F.~Wirth, N.~Lo~Iudice,
                 A.~V.~Sushkov, N.~Yu.~Shirikova,  Y.~Sun, T.~Faestermann, R.~Kr\"{u}cken,
                 M.~Mahgoub, J.~Jolie, P.~von~Brentano, N.~Braun, S.~Heinze, O.~M\"{o}ller,
                 D.~M\"{u}cher, C.~Scholl, R.F.~Casten, D.A.~Meyer,
                 Phys. Rev. C {\bf 73}, 064309 (2006).

\bibitem{Bet09}  L.~Bettermann, S.~Heinze, J.~Jolie, D.~M\"{u}cher, O.~M\"{o}ller, C.~Scholl,
                 R.~F.~Casten, D.~A.~Meyer, G. Graw, R. Hertenberger, H.-F.~Wirth, and D.~Bucurescu,
                 Phys. Rev. C {\bf 80}, 044333 (2009).

\bibitem{Ili10}  G.~Ilie, R.~F.~Casten, P.~von~Brentano, D.~Bucurescu, T.~Faestermann,
                 G.~Graw, S.~Heinze, R.~Hertenberger, J.~Jolie,  R.~Kr\"{u}cken, D.~A.~Meyer,
                 D.~M\"{u}cher, C.~Scholl, V.~Werner, R.~Winkler, and H.-F.~Wirth,
                 Phys. Rev. C {\bf 82}, 024303 (2010).

 \bibitem{Ber13} C.~Bernards, R.~F.~Casten, V.~Werner, P.~von~Brentano, D.~Bucurescu, G.~Graw,
                 S.~Heinze, R.~Hertenberger, J.~Jolie, S.~Lalkovski, D.~A.~Meyer, D.~M\"{u}cher,
                 P.~Pejovic, C.~Scholl, and H.-F.~Wirth, Phys. Rev. C {\bf 87}, 024318 (2013).

\bibitem{AME16}  Atomic Mass Evaluation - AME2016, Chiness Physics C {\bf41}
                  2017 030002.

\bibitem{Rie91} F.~Riess, Beschleunigerlaboratorium M\"unchen, Annual
                          Report, 1991, p.168.

\bibitem{Kun} P.~D.~Kunz, computer code CHUCK3, University of Colorado, unpublished.

\bibitem{Bec69} F.~D.~Becchetti and G.~W.~Greenlees, Phys. Rev. {\bf 182},  1190 (1969).

\bibitem{Fly69} E.~R.~Flynn, D.~D.~Amstrong, J.~G.~Beery, and A.~G.~Blair,
                Phys. Rev. {\bf 182}, 1113 (1969).

\bibitem{Gre78}  R.~C.~Greenwood, C.~W.~Reich, H.~A.~Baader, H.~R.~Koch, D.~Breitig,
                 O.~W.~B.~Schult, B.~Fogelberg, A.~Backlin, W.~Mampe, T.~von~Egidy,
                 E.~Schreckenbach, Nucl. Phys. A{\bf304}, 327 (1978).

\bibitem{Lesh07} S.~R.~Lesher, J.~N.~Orce, Z.~Ammar, C.~D.~Hannant, M.~Merrick, N.~Warr,
                 T.~B.~Brown,
                N.~Boukharouba, C.~Fransen, M.~Scheck, M.~T.~McEllistrem, and S.~W.~Yates,
                Phys. Rev C{\bf 76}, 034318 (2007)

\bibitem{Nic17}  N.~Nica, Nucl. Data Sheets 141, 1 (2017).

\bibitem{Pas10}  S.~Pascu, N.~V.~Zamfir, Gh.~Cata-Danil, and N.~Marginean,
                 Phys. Rev. C{\bf81}, 054321 (2010).

\bibitem{Lev18}  A.~I.~Levon, A.~G.~Magner, and
                S.~V.~Radionov, Phys.Rev. C  {\bf 97}, 044305 (2018).

\bibitem{Mag18}   A.~G.~Magner, A.~I.~Levon, and S.~V.~Radionov,
arXiv preprint arXiv:1804.01433;  Eur. J. Phys., in press (2018).

\end{thebibliography}
\end{document}